\documentclass{article}

 \usepackage[preprint]{neurips_2026}

\usepackage[utf8]{inputenc} % allow utf-8 input
\usepackage[T1]{fontenc}    % use 8-bit T1 fonts
\usepackage{hyperref}       % hyperlinks
\usepackage{url}            % simple URL typesetting
\usepackage{booktabs}       % professional-quality tables
\usepackage{amsfonts}       % blackboard math symbols
\usepackage{nicefrac}       % compact symbols for 1/2, etc.
\usepackage{microtype}      % microtypography
\usepackage{xcolor}         % colors
\usepackage{graphicx}
\usepackage{amsmath}
\usepackage{bbm}
\usepackage{wrapfig}
\usepackage[table]{xcolor}

\title{Sequence-Informed Geometric Evaluation of RNA 3D Structures}

\author{%
  Andrea Zerio\textsuperscript{1,2}
  \And
  Yighua Yao\textsuperscript{2}
  \And
  Alessandro Micheli\textsuperscript{5}
  \And
  Roland G. Huber\textsuperscript{4}
  \AND
  Mile Sikic\textsuperscript{3}
  \And
  Samir Bhatt\textsuperscript{5,6,7}
  \And
  Andres R. Masegosa\textsuperscript{1}
  \And
  Yuangang Pan\textsuperscript{2}
  \\[1.5ex]
  \textsuperscript{1}Department of Computer Science, Aalborg University,
  Copenhagen, Denmark \\
  \textsuperscript{2}Centre for Frontier AI Research (CFAR),\\
  Institute of Advanced Intelligence and Computing (IAIC),
  A*STAR, Singapore \\
  \textsuperscript{3}Genome Institute of Singapore (GIS),
  A*STAR, Singapore \\
  \textsuperscript{4}Bioinformatics Institute (BII),
  A*STAR, Singapore \\
  \textsuperscript{5}Imperial College London, London, UK \\
  \textsuperscript{6}Statens Serum Institut, Copenhagen, Denmark \\
  \textsuperscript{7}University of Copenhagen, Copenhagen, Denmark \\
}

\begin{document}

\maketitle

\begin{abstract}
Computational RNA structure pipelines generate many candidate conformations for the same sequence. Reliable evaluation therefore requires more than recognising
plausible geometry, it requires determining whether that geometry is compatible with the sequence. We introduce SIRGE, a sequence-informed geometric evaluator that conditions structural representations on nucleotide embeddings from a pretrained RNA language model. Early results show that SIRGE outperforms established evaluators in Kendall--$\tau$ alignment, Top-1 selection, and Top-3 ranking. Controlled comparisons further show that sequence conditioning corrects errors made by an otherwise matched geometric model and improves target-level rank structure. These findings provide initial evidence that pretrained sequence representations supply ranking information that complements geometric reasoning.
\end{abstract}
\section{Introduction}
\label{sec:introduction}

RNA molecules regulate gene expression, catalyse biochemical reactions, and are an expanding class of therapeutic targets \cite{cech2014noncoding}. Their functions depend critically on three-dimensional structure \cite{churkin2018design, ponce2019computational}, yet determining RNA conformations remains difficult and costly \cite{renaud2018cryo, jin2025computational}. Computational pipelines therefore often produce multiple candidates of varying quality for the same sequence \cite{li2025homrank, tarafder2024lociparse}, making reliable evaluation essential, since errors at this stage can negate gains made by the predictor itself \citep{townshend2021geometric,li2025equirna}. RNA structure evaluation is typically approached primarily as a geometric problem. Given a candidate structure, an evaluator reasons over its atoms, distances, orientations, and higher-order geometric features to estimate its quality. We argue that this view is incomplete and that candidate evaluation is fundamentally a problem of \textbf{sequence--geometry compatibility}. Our argument is that the quality of a candidate is not determined by its geometry in isolation, but by whether that geometry is compatible with the particular RNA sequence that is expected to adopt it. Geometry describes the interactions realised by a candidate conformation, while the sequence provides context about which structural arrangements are plausible for that RNA. Because the sequence is readily available when candidates are scored, learning this sequence–geometry alignment may also improve generalisation: pretrained sequence representations can provide transferable priors for evaluating RNA targets and candidate conformations not seen during supervised training.

Recent methods have made geometric reasoning for RNA evaluation increasingly expressive \citep{liu2026rnarank, li2025homrank, townshend2021geometric, tarafder2024lociparse, li2025equirna}, but they do not condition these geometric representations on contextual nucleotide embeddings pretrained across large RNA sequence collections. RNA language models~\citep{chen2022interpretable,abramson2024accurate,penic2025rinalmo} provide a natural source of such information. Through self-supervised pretraining, they learn nucleotide representations that capture dependencies across RNA sequences and transfer to structure-related tasks. RiNALMo, for example, learns contextual nucleotide embeddings from millions of non-coding RNA sequences and demonstrates strong transfer to RNA structure prediction tasks \citep{penic2025rinalmo}. This raises a natural question: \emph{can sequence context help identify the best RNA structure candidate?}

\begin{figure}[t]
    \centering
    % Replace the path below with the final architecture figure.
    \includegraphics[width=\linewidth]{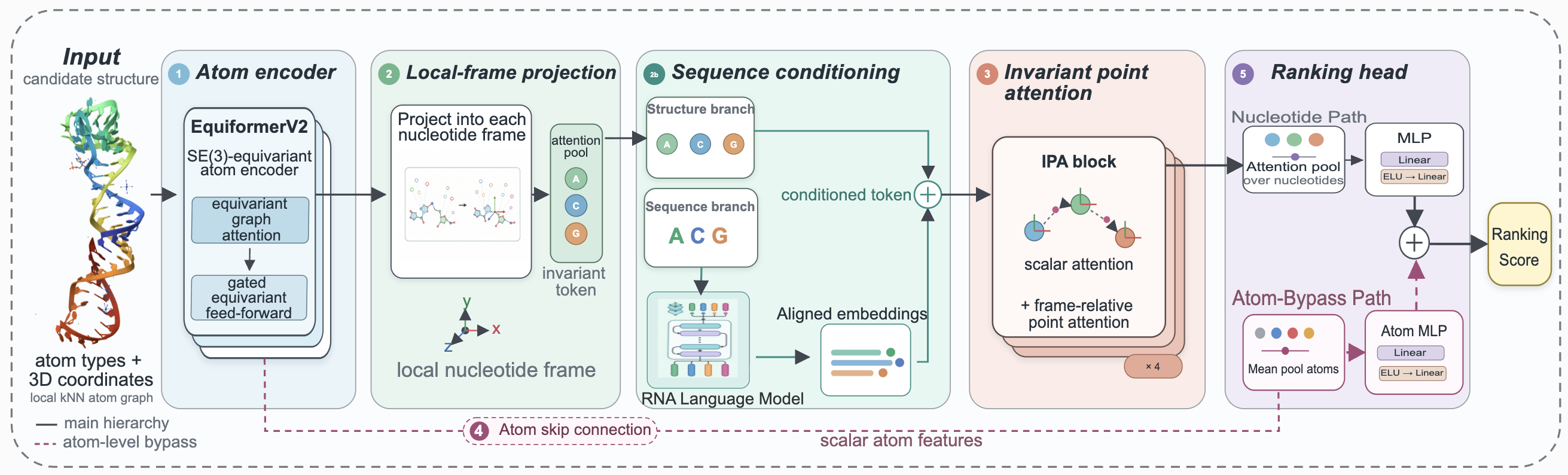}
    \caption{
        \textbf{Overview of SIRGE.}
        An $\mathrm{SE}(3)$-equivariant atom transformer encodes the candidate structure.
        Atom representations are projected into nucleotide-local frames and combined with contextual embeddings from a pretrained RNA language model.
        Dense invariant point attention integrates information across nucleotides, while atom residual connections preserve fine-grained geometry for the final scoring head.
    }
    \label{fig:sirge_architecture}
\end{figure}

We introduce \textbf{SIRGE}, a \textbf{S}equence-\textbf{I}nformed \textbf{R}NA \textbf{G}eometric \textbf{E}valuator for ranking 3D structure candidates.  As illustrated in Fig.~\ref{fig:sirge_architecture}, SIRGE first encodes the candidate at atomic resolution with an $\mathrm{SE}(3)$-equivariant atom transformer from the Equiformer family \citep{liao2024equiformerv2} and then projects the resulting atom representations into nucleotide-local frames \citep{tarafder2024lociparse}. SIRGE combines each structure-derived nucleotide representation with its corresponding language-model embedding. Dense invariant point attention propagates information across the nucleotide frames \citep{jumper2021applying,tarafder2024lociparse}. Finally, atom-level residual connections restore fine-grained geometric information before a predictive head produces the candidate score. We compare SIRGE with established RNA structure evaluation baselines and a structure-only variant of our architecture. Our early results show substantial improvements in Kendall--$\tau$ rank alignment, top-1 candidate retrieval, and top-3 candidate ranking. Pairwise and target-level analyses further illustrate how sequence-informed geometric reasoning improves the ordering of candidate structures.

Our main \textbf{contributions} are threefold: \textbf{(i)} We formulate RNA 3D structure evaluation as sequence-conditioned geometric ranking. This perspective treats candidate quality as compatibility between an RNA sequence and a proposed three-dimensional conformation. \textbf{(ii)} We introduce SIRGE, a sequence-informed geometric evaluator that conditions candidate structure representations on contextual embeddings from a pretrained RNA language model. \textbf{(iii)} We provide early empirical evidence that sequence conditioning improves RNA candidate evaluation across complementary ranking and retrieval metrics.

\section{Sequence-Informed Geometric Evaluation}
\label{sec:method}

Following HomRank, we train SIRGE using homogeneous batches of candidate structures \(X_m\) for a single RNA sequence \(S=(s_i)_{i=1}^{T}\), where \(m,n\) index candidates and \(i,j\) index nucleotides. Let \(d_m\) denote the RMSD of \(X_m\) from the native structure. SIRGE learns a scoring function \(f_\theta(X_m,S)\in\mathbb{R}\) whose ordering is the inverse of the RMSD ordering:
\begin{equation}
    d_m < d_n
    \quad \Longrightarrow \quad
    f_\theta(X_m,S) > f_\theta(X_n,S).
    \label{eq:ranking_goal}
\end{equation}
Within each batch, we sample candidate lists and optimise their ordering using the ListNet objective \citep{cao2007learning}; its full definition is provided in Appendix~\ref{app:ranking_objective}. Because the supervision depends on within-RNA ordering rather than absolute RMSD, it is invariant to sequence length and provides a less noisy training signal. Additionally, and unlike RMSD, $f_\theta$ can be evaluated without the native structure, using only the candidate coordinates and RNA sequence.

\paragraph{Hierarchical geometric representation}
\label{sec:geometric_representation}

Candidate $X_m$ contains $N_m$ atoms. Each atom
$a\in\{1,\ldots,N_m\}$ has coordinate
$\mathbf{x}_{m,a}\in\mathbb{R}^3$, element one-hot
$\mathbf{c}_{m,a}\in\{0,1\}^5$ and nucleotide assignment $\rho(a)\in\{1,\ldots,T\}$. An Equiformer atom encoder
$\Phi_{\mathrm{atom}}$ \citep{liao2024equiformerv2} maps the atom graph to
equivariant features:
\begin{equation}
\left\{\mathbf{a}_{m,a}^{(\ell)}\right\}_{a,\ell}
=
\Phi_{\mathrm{atom}}
\left(
\left\{(\mathbf{c}_{m,a},\mathbf{x}_{m,a})\right\}_{a=1}^{N_m}
\right),
\qquad
\ell=0,\ldots,\ell_{\max}.
\label{eq:atom_encoder}
\end{equation}
Here, $\ell=0$ denotes rotation-invariant scalar channels, $\ell=1$ denotes
vector and $\ell>1$ denotes higher-order $\mathrm{SO}(3)$ representations. Following AlphaFold2 and lociPARSE \citep{jumper2021applying,tarafder2024lociparse}, each nucleotide
$i$ defines a local frame
$F_{m,i}=(\mathbf{R}_{m,i},\mathbf{o}_{m,i})$, where
$\mathbf{R}_{m,i}\in\mathrm{SO}(3)$ specifies its axes and
$\mathbf{o}_{m,i}\in\mathbb{R}^3$ its origin. For each atom $a$ with
$\rho(a)=i$, SIRGE constructs
\begin{equation}
\mathbf{d}_{m,a}
=
\left[
\mathbf{a}_{m,a}^{(0)},
\mathbf{R}_{m,i}^{\top}\mathbf{a}_{m,a}^{(1)},
\mathbf{R}_{m,i}^{\top}
\left(\mathbf{x}_{m,a}-\mathbf{o}_{m,i}\right)
\right].
\label{eq:local_projection}
\end{equation}
The brackets denote concatenation, and multiplication by
$\mathbf{R}_{m,i}^{\top}$ expresses the vector channels and centred coordinates
in the nucleotide frame. Because the local frame co-transforms with the candidate, $\mathbf{d}_{m,a}$ is invariant to global $\mathrm{SE}(3)$ transformations; Appendix~\ref{app:rigid_invariance} provides the
full argument. Attention pooling then maps
$\{\mathbf{d}_{m,a}:\rho(a)=i\}$ to the invariant geometric representation $\widetilde{\mathbf{u}}_{m,i}$ of nucleotide $i$ in candidate $X_m$.

\paragraph{Sequence--geometry conditioning}
\label{sec:sequence_conditioning}

RNA sequences constrain which spatial interactions can stabilise a fold.
These constraints extend beyond individual base identities: substitutions
that preserve Watson--Crick pairing can alter helix conformational
preferences and tertiary stability \citep{yesselman2019sequence}, while
energetic effects within tertiary motifs depend on interactions among
their constituent residues \citep{shin2023dissecting}. A candidate can
therefore satisfy general geometric regularities without reproducing the
structural organisation favoured by its sequence. We frame evaluation
as sequence--geometry alignment: assessing the compatibility of the
geometry realised by $X_m$ with folding constraints associated with $S$. 

We first augment each pooled geometric representation with nucleotide
identity. Let $\mathbf{r}_i=\operatorname{onehot}(s_i)\in\{0,1\}^{4}$,
with entries corresponding to A, C, G, and U, and define
$\mathbf{u}_{m,i}=[\widetilde{\mathbf{u}}_{m,i},\mathbf{r}_i]$.
The identity vector assigns the same encoding to identical nucleotides
regardless of their surrounding sequence, even though their structural roles
can depend on both nearby nucleotides and long-range sequence
relationships. To represent these dependencies, we use a pretrained RNA
sequence encoder $\psi$ that produces one contextual embedding per
nucleotide:
$\mathbf{E}=\psi(S)=(\mathbf{e}_1,\ldots,\mathbf{e}_T)$.
Each $\mathbf{e}_i$ describes nucleotide $i$ in the context of the
complete sequence.

In our implementation, $\psi$ is RiNALMo \citep{penic2025rinalmo},
pretrained by masked language modelling on 36 million ncRNA sequences,
and $\mathbf{e}_i\in\mathbb{R}^{1280}$ is its final-layer representation.
Predicting masked nucleotides encourages the encoder to capture sequence
dependencies that can reflect conserved pairing patterns and structural
motifs. RiNALMo's transfer to secondary-structure
prediction on RNA families withheld from downstream training motivates
its use as a source of folding-related sequence features. We keep $\psi$
frozen to preserve this pretrained representation. SIRGE incorporates each contextual embedding into the representation of
the corresponding nucleotide through a learned affine projection:
\begin{equation}
\mathbf{h}_{m,i}^{\mathrm{cond}}
=
\mathbf{u}_{m,i}
+
\mathbf{W}_{E}\operatorname{LN}(\mathbf{e}_i)
+
\boldsymbol{\beta}_{E}.
\label{eq:sequence_conditioning}
\end{equation}
Here, $\operatorname{LN}$ denotes layer normalisation, while
$\mathbf{W}_{E}$ and $\boldsymbol{\beta}_{E}$ are the weight matrix
and bias of an affine map from $\mathbb{R}^{1280}$ to the representation
space of $\mathbf{u}_{m,i}$. Both are initialised to zero, so
$\mathbf{h}_{m,i}^{\mathrm{cond}}=\mathbf{u}_{m,i}$ at initialisation. Conditioning at corresponding nucleotide indices allows the evaluator
to assess whether the spatial environment proposed in each candidate
agrees with the structural preferences reflected in its sequence context.
The geometric representations are invariant to global rigid motions,
so this assessment concerns the candidate's internal spatial organisation.

All candidates of an RNA share $\mathbf{E}$, but their geometries can
differ in their compatibility with that context. A sequence-only term
added to the final score would instead shift every candidate equally,
leaving both the ordering and the ListNet objective unchanged
(Appendix~\ref{app:sequence_ranking}). Because nucleotide identities already specify $S$, the contribution of
$\mathbf{E}=\psi(S)$ is a pretrained representation of existing sequence
information. Its dependencies are learned from a much larger sequence
corpus than the set of RNAs available for supervised ranking, making
this knowledge available without requiring the evaluator to discover
it from ranked structures alone. Our claim is that learning
sequence--geometry compatibility with this transferable context improves candidate ranking and generalisation to unseen RNAs.

\paragraph{Geometric reasoning and readout}
\label{sec:ipa_readout}

Four dense invariant point attention (IPA) blocks \citep{jumper2021applying} update the conditioned nucleotide representations. At block $b$, head $r$ projects $\mathbf{h}_{m,i}^{[b]}$ into scalar queries
and keys, $\mathbf{q}_{m,i}^{[b,r]}$ and $\mathbf{k}_{m,i}^{[b,r]}$, and $P$
learned 3D points, $\widehat{\mathbf{q}}_{m,i,p}^{[b,r]}$ and
$\widehat{\mathbf{k}}_{m,i,p}^{[b,r]}$. Their attention logit is
\begin{equation}
\begin{split}
s_{m,i,j}^{[b,r]}
&=
\frac{
\left\langle
\mathbf{q}_{m,i}^{[b,r]},
\mathbf{k}_{m,j}^{[b,r]}
\right\rangle
}{\sqrt{D}}
\quad-
\frac{\gamma_{b,r}}{2}
\sum_{p=1}^{P}
\left\|
\mathcal{T}_{m,i}
\left(\widehat{\mathbf{q}}_{m,i,p}^{[b,r]}\right)
-
\mathcal{T}_{m,j}
\left(\widehat{\mathbf{k}}_{m,j,p}^{[b,r]}\right)
\right\|_2^2 .
\label{eq:ipa_attention}
\end{split}
\end{equation}
where $D$ is the scalar head width, $\gamma_{b,r}>0$ is a trainable per-head weight controlling the strength of the geometric distance penalty, and $\mathcal{T}_{m,i}(\mathbf{z})=\mathbf{R}_{m,i}\mathbf{z}+\mathbf{o}_{m,i}$
maps a local point into global coordinates. Softmax-normalised logits weight learned value projections to produce $\mathbf{h}_{m,i}^{[b+1]}$. Because the
point distances are unchanged by a global rigid transformation, IPA captures long-range interactions while preserving $\mathrm{SE}(3)$ invariance. After four blocks, attention pooling produces the candidate representation $\mathbf{g}_m$. A residual atom path applies a learned projection $\eta_0$ to the Equiformer scalar features and mean-pools them. The two paths produce the final score:
\begin{equation}
f_\theta(X_m,S)
=
\phi_{\mathrm{nt}}(\mathbf{g}_m)
+
\phi_{\mathrm{atom}}
\Big(
\frac{1}{N_m}
\sum_{a=1}^{N_m}
\eta_0\!\big(\mathbf{a}_{m,a}^{(0)}\big)
\Big),
\label{eq:final_score}
\end{equation}
where $\eta_0$ is a learned atom-level projection and
$\phi_{\mathrm{nt}}$ and $\phi_{\mathrm{atom}}$ are multilayer perceptrons.
The residual path preserves invariant atom-level information that may be
diluted during atom-to-nucleotide aggregation.

\section{Experimental Validation}
\label{sec:experiments}

\paragraph{Dataset and experimental setup}
\label{sec:experimental_setup}

% \begin{wrapfigure}{r}{0.55\linewidth}
% % \vskip-0.75in
% \centering
% \includegraphics[width=\linewidth]{figures/fig1b_pairwise_accuracy_gain.png}
% % \vskip-0.1in
% \caption{\label{fig:pairwise_evidence} Pairwise-accuracy change after adding RiNALMo conditioning to a matched structure-only backbone. Candidate quality is expressed as within-RNA RMSD percentile, where $0$ is best. Blue indicates a gain from conditioning.}
% % \vskip-0.5in
% \end{wrapfigure}

\begin{figure}[t]
\centering
\includegraphics[width=0.5\linewidth]{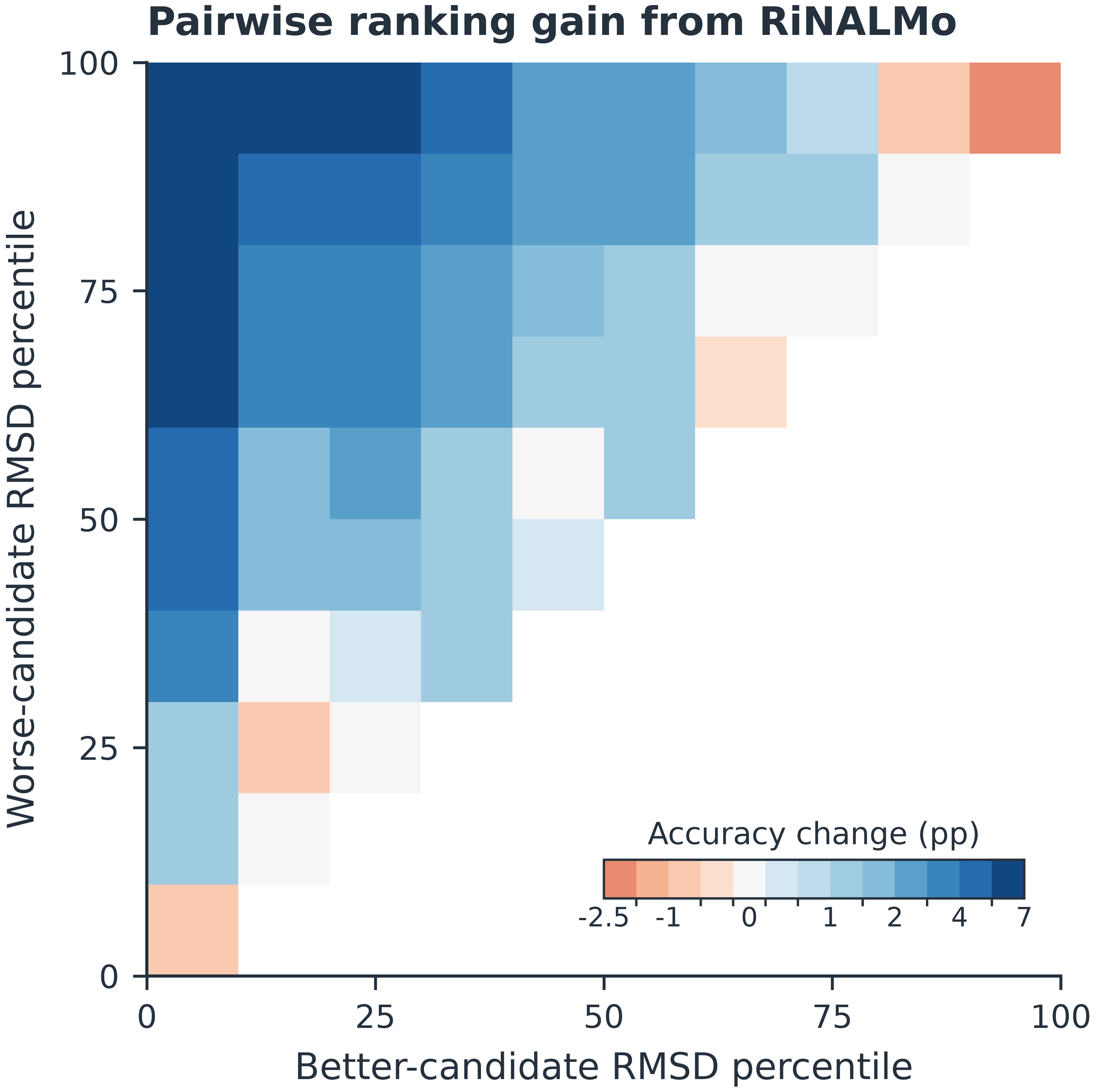}
\caption{
Pairwise-accuracy change after adding RiNALMo conditioning to a matched
structure-only backbone. Candidate quality is expressed as within-RNA RMSD
percentile, where $0$ is best. Blue indicates a gain from conditioning.
}
\label{fig:pairwise_evidence}
\end{figure}
% \begin{wraptable}{r}{0.75\linewidth}
% % \vskip-0.2in
% \centering
% \caption{\label{tab:main_results} Global and band-specific Kendall--$\tau$ rank correlation on \mbox{Test-CLS-1}. Results are reported as mean $\pm$ std over three random seeds.}
% \scriptsize
% \setlength{\tabcolsep}{2.5pt}
% \renewcommand{\arraystretch}{0.92}
% \resizebox{\linewidth}{!}{
% \begin{tabular}{@{}lcccc@{}}
% \toprule
% Model
% & Global
% & D1 ($2$--$5$\,\AA)
% & D2 ($5$--$10$\,\AA)
% & D3 ($10$--$15$\,\AA) \\
% \midrule

% Rosetta & 0.016 & 0.005 & -0.024 & 0.008 \\

% ARES 
% & 0.130 $\pm$ 0.013 
% & 0.218 $\pm$ 0.018 
% & 0.153 $\pm$ 0.009 
% & 0.064 $\pm$ 0.006 \\

% PAMNet
% & $0.068 \pm 0.003$
% & $0.166 \pm 0.016$
% & $0.047 \pm 0.008$
% & $-0.033 \pm 0.014$ \\

% RNA3DCNN
% & $0.074 \pm 0.014$
% & $0.138 \pm 0.008$
% & $0.084 \pm 0.023$
% & $0.020 \pm 0.006$ \\

% EquiRNA
% & $0.075 \pm 0.016$
% & $0.173 \pm 0.056$
% & $0.088 \pm 0.024$
% & $0.057 \pm 0.016$ \\

% lociPARSE
% & $0.135 \pm 0.010$
% & $0.261 \pm 0.014$
% & $0.140 \pm 0.018$
% & $0.060 \pm 0.025$ \\

% HomRank
% & $0.146 \pm 0.012$
% & $0.231 \pm 0.026$
% & $0.208 \pm 0.018$
% & $0.109 \pm 0.013$ \\

% \rowcolor{gray!25}
% \textbf{SIRGE-SO}
% & $0.209 \pm \mathrm{0.010}$
% & $0.381 \pm \mathrm{0.003}$
% & $0.282 \pm \mathrm{0.020}$
% & $0.148 \pm \mathrm{0.001}$ \\

% \rowcolor{gray!25}
% \textbf{SIRGE}
% & $\mathbf{0.255} \pm \mathrm{0.008}$
% & $\mathbf{0.430} \pm \mathrm{0.001}$
% & $\mathbf{0.341} \pm \mathrm{0.001}$
% & $\mathbf{0.204} \pm \mathrm{0.017}$ \\
% \bottomrule
% \end{tabular}
% }
% % \vspace{-1.2em}
% \end{wraptable}
\begin{table}[t]
\centering
\caption{
Global and band-specific Kendall--$\tau$ rank correlation on
\mbox{Test-CLS-1}. Results are reported as mean $\pm$ std over three random seeds.
}
\label{tab:main_results}

\scriptsize
\setlength{\tabcolsep}{4pt}
\renewcommand{\arraystretch}{1.18}

\resizebox{\linewidth}{!}{
\begin{tabular}{@{}lcccc@{}}
\toprule
Model
& Global
& D1 ($2$--$5$\,\AA)
& D2 ($5$--$10$\,\AA)
& D3 ($10$--$15$\,\AA) \\
\midrule

Rosetta
& 0.016
& 0.005
& -0.024
& 0.008 \\[2pt]

ARES
& $0.130 \pm 0.013$
& $0.218 \pm 0.018$
& $0.153 \pm 0.009$
& $0.064 \pm 0.006$ \\

PAMNet
& $0.068 \pm 0.003$
& $0.166 \pm 0.016$
& $0.047 \pm 0.008$
& $-0.033 \pm 0.014$ \\

RNA3DCNN
& $0.074 \pm 0.014$
& $0.138 \pm 0.008$
& $0.084 \pm 0.023$
& $0.020 \pm 0.006$ \\

EquiRNA
& $0.075 \pm 0.016$
& $0.173 \pm 0.056$
& $0.088 \pm 0.024$
& $0.057 \pm 0.016$ \\

lociPARSE
& $0.135 \pm 0.010$
& $0.261 \pm 0.014$
& $0.140 \pm 0.018$
& $0.060 \pm 0.025$ \\[2pt]

HomRank
& $0.146 \pm 0.012$
& $0.231 \pm 0.026$
& $0.208 \pm 0.018$
& $0.109 \pm 0.013$ \\

\midrule

\rowcolor{gray!20}
\textbf{SIRGE-SO}
& $0.209 \pm 0.010$
& $0.381 \pm 0.003$
& $0.282 \pm 0.020$
& $0.148 \pm 0.001$ \\

\rowcolor{gray!20}
\textbf{SIRGE}
& $\mathbf{0.255} \pm 0.008$
& $\mathbf{0.430} \pm 0.001$
& $\mathbf{0.341} \pm 0.001$
& $\mathbf{0.204} \pm 0.017$ \\

\bottomrule
\end{tabular}
}
\end{table}

We use the RNA structure-evaluation benchmark introduced by HomRank \citep{li2025homrank}, comprising 190 non-redundant single-chain RNA targets drawn from experimental structures and established benchmarks. For newly collected references, HomRank generates 1,000 candidate conformations using molecular dynamics and supplements these with candidate pools from ARES and RNA-Puzzles. Full construction details are provided in Appendix~\ref{app:dataset}. We compare SIRGE with 7 baselines as well as SIRGE-SO, the same architecture without RiNALMo conditioning, to separate architectural gains from the contribution of sequence information. All neural models are trained for five epochs, selected on the validation set, and evaluated on Test-CLS-1. Further details appear in Appendix~\ref{app:experimental_protocol}.

We report signed Kendall--$\tau$ alignment, $\tau_{\mathrm{align}}=\tau(f_\theta,-d)$, where $+1$ denotes perfect agreement, $0$ no rank association, and $-1$ a completely inverted ordering. We report global alignment over the full candidate pool and stratify uniformly sampled lists of ten candidates by the RMSD of their best structure: D1 ($2$--$5\,\text{\AA}$), D2 ($5$--$10\,\text{\AA}$), and D3 ($10$--$15\,\text{\AA}$). We focus on these bands as they represent the quality range most relevant to selecting among computationally generated RNA structures \citep{li2025homrank}. We further report Top-1 retrieval, Top-3 ranking, and Spearman  correlation in Appendix~\ref{app:complete_results}.

\begin{figure*}[t]
\centering

\makebox[\textwidth][c]{%
    \includegraphics[width=1.06\textwidth]{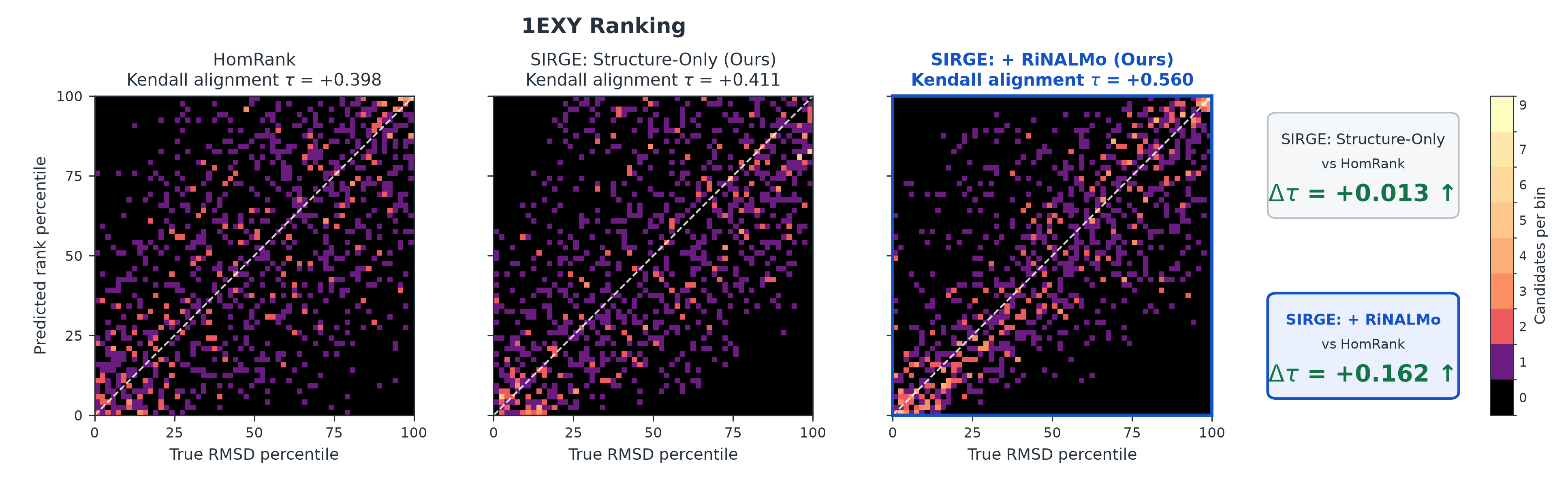}
}

\vspace{0.3em}

\makebox[\textwidth][c]{%
    \includegraphics[width=1.06\textwidth]{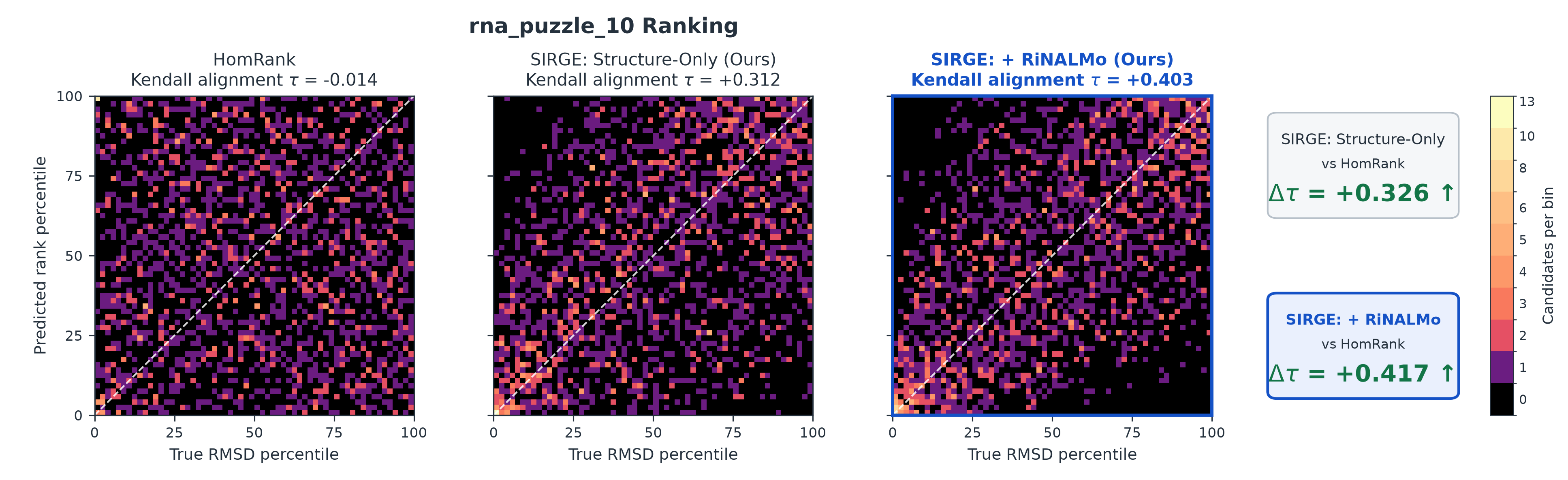}
}

\caption{
\textbf{Sequence conditioning improves target-level candidate ordering.}
Rank-density maps for two RNAs comparing HomRank, SIRGE-SO, and SIRGE.
Candidates are placed by true RMSD percentile (x-axis) and predicted rank
percentile (y-axis), with $0$ best on both axes. Colour shows candidate
density per bin. Density near the dashed diagonal indicates rank agreement;
off-diagonal density indicates ranking error. SIRGE shows a sharper diagonal
concentration and fewer dispersed high-density regions, indicating more
coherent ordering from sequence conditioning.
}
\label{fig:rank_density}
\end{figure*}

% \begin{figure*}[t]
% % \vspace{-1.8em}
% \centering
% \includegraphics[width=\textwidth]{figures/1EXY_bins60.png}

% \includegraphics[width=\textwidth]{figures/rna_puzzle_10_bins60.png}
% \caption{
% \textbf{Sequence conditioning improves target-level candidate ordering.}
% Rank-density maps for two RNAs comparing HomRank, SIRGE-SO, and SIRGE. Candidates are placed by true RMSD percentile (x-axis) and predicted rank percentile (y-axis), with $0$ best on both axes. Colour shows candidate density per bin. Density near the dashed diagonal indicates rank agreement; off-diagonal density indicates ranking error. SIRGE shows a sharper diagonal concentration and fewer dispersed high-density regions, indicating more coherent ordering from sequence conditioning.
% }
% \label{fig:rank_density}
% % \vspace{-1.5em}
% \end{figure*}

\paragraph{Sequence information complements geometry}
\label{sec:main_results}

Table~\ref{tab:main_results} shows that SIRGE achieves the strongest global and band-specific rank alignment, with  SIRGE-so ablation results separating the two sources of this improvement. While the structure-only variant already outperforms previous geometric evaluators, adding sequence-conditioning produces further gains in every quality band. Moreover these gains become more pronounced as candidate quality decreases, indicating that sequence context is especially valuable when the proposed geometry is less consistent with the RNA. This pattern suggests that the sequence embeddings encode implicit conformational priors that complement geometric features. 
\\

Crucially, all candidates for a given RNA share the same sequence, so a sequence-only score cannot alter their relative ordering. SIRGE instead uses sequence context to interpret each candidate geometry. The pairwise analysis in  Figure~\ref{fig:pairwise_evidence}
corroborates this interpretation, with the largest gains involving poorer candidate structures, while figure~\ref{fig:rank_density} provides a target-level view of this improvement. Bright density along the diagonal indicates consistent agreement between predicted and true ranks, while diffuse density reflects noisy predictions and
bright off-diagonal or anti-diagonal structure reveals systematic or inverted rankings. Conditioning concentrates the candidate density closer to the diagonal, bringing predicted ranks into clearer agreement with the true
ordering. 
\section{Conclusion}
\label{sec:conclusion}

We introduced SIRGE, an RNA structure evaluator that treats candidate ranking as a problem of sequence--geometry compatibility. Our early results show that conditioning geometric representations on pretrained RNA language-model
embeddings improves global rank alignment and the selection of high-quality candidates. Controlled analyses provide further evidence that sequence information complements the signal available from geometry alone. We next plan to investigate richer forms of sequence--structure conditioning and test SIRGE on candidates generated by modern structure-prediction models.

\newpage

\bibliographystyle{plainnat}
\bibliography{references}

%%%%%%%%%%%%%%%%%%%%%%%%%%%%%%%%%%%%%%%%%%%%%%%%%%%%%%%%%%%%
\newpage
\appendix
% Early evaluation methods used knowledge-based statistical potentials. For example, 3dRNAscore combines distance- and torsion-dependent terms derived from experimentally determined structures \citep{wang2015_3drnascore}. Learned evaluators later replaced hand-crafted potentials with representations learned directly from structural data. RNA3DCNN applies three-dimensional convolutions to voxelised atomic environments \citep{li2018_rna3dcnn}, while ARES uses rotationally equivariant message passing over atom types and coordinates \citep{townshend2021geometric}. More recent architectures capture structure at multiple levels. lociPARSE represents nucleotides through local coordinate frames and uses invariant point attention to estimate structural quality \citep{tarafder2024_lociparse}. EquiRNA combines atom-, nucleotide-, and RNA-level geometries to improve generalisation across RNA sizes \citep{li2025equirna}. RNArank instead integrates engineered one-, two-, and three-dimensional features and predicts intermediate contact and distance-deviation maps \citep{liu2026rnarank}. HomRank further studies structure evaluation directly as a within-target ranking problem \citep{li2025homrank}.

\section{Related Work}
\label{app:related_work}

Early RNA structure evaluators relied on hand-crafted objectives, either combining knowledge-based potentials with physics-inspired energy terms \citep{xu2015physics} or deriving distance- and torsion-dependent potentials
from experimental structures \citep{wang20153drnascore}. However their accuracy is largely limited by how well hand-crafted terms approximate RNA folding physics \cite{tan2019best, li2025homrank}

Learned evaluators instead derive representations directly from candidate geometry. RNA3DCNN applies convolutions to voxelised atomic environments \citep{li2018rna3dcnn}, while ARES and PAMNet use geometric graph networks \citep{townshend2021geometric,zhang2023universal}. More recent methods introduce richer structural organisation. lociPARSE uses nucleotide-local frames and invariant point attention \citep{tarafder2024lociparse}; EquiRNA models RNA
through hierarchical equivariant representations \citep{li2025equirna}; and RNArank predicts intermediate contact and distance-deviation maps \citep{liu2026rnarank}. HomRank further formulates evaluation as within-RNA
learning to rank \citep{li2025homrank}. These methods improve geometric reasoning but do not condition it on contextual representations learned through large-scale RNA sequence pretraining.

RNA language models provide such representations. RNA-FM \citep{chen2022interpretable} and RiNALMo \citep{penic2025rinalmo} learn contextual nucleotide embeddings from unlabelled RNA sequences and transfer them to structural and functional tasks . Their use in structure modelling has focused primarily on predicting structural properties rather than ranking alternative conformations. SIRGE connects these lines of work by using pretrained nucleotide embeddings to condition a symmetry-aware geometric evaluator.
\section{Formal Properties and Learning Objective}
\label{app:formal_details}

\subsection{Sequence--geometry alignment}
\label{app:sequence_ranking}

\paragraph{Structural compatibility.}
Sequence--geometry alignment denotes a learned assessment of whether a
candidate's spatial organisation agrees with the folding preferences
associated with its sequence. These preferences reflect
sequence-dependent effects on helix conformations and tertiary
interactions \citep{yesselman2019sequence,shin2023dissecting}.
The geometric representation describes the environment realised by a
candidate, while the contextual sequence representation provides
pretrained features from which structural preferences can be inferred.
Their correspondence at nucleotide index $i$ allows the evaluator to
relate these descriptions of the same physical nucleotide.

This correspondence is useful because the two representations respond
differently to errors in a proposed structure. The atom encoder constructs
its neighbourhoods from candidate coordinates, so changes in geometry
can change both the spatial interactions represented and the information
exchanged between atoms. In contrast, $\mathbf{e}_i$ depends only on
$S$ and remains fixed across alternative conformations. It therefore
provides a consistent sequence context against which their differing
geometries can be assessed. Both the projected geometric features and
the sequence embeddings are invariant to global rigid motions, so this
assessment concerns internal spatial organisation
(Appendix~\ref{app:rigid_invariance}). The ranking objective learns how
this compatibility relates to the reference RMSD ordering.

\paragraph{The shared sequence alone cannot change ranking.}
Consider a list $L$ of candidate indices for a fixed RNA sequence $S$,
with shared embeddings $\mathbf{E}=\psi(S)$. Let
$\kappa(\mathbf{E})\in\mathbb{R}$ be any scalar computed solely from
these embeddings. To isolate the effect of adding this sequence-only
contribution, define the auxiliary score
\begin{equation}
f_\theta^{+}(X_m,S)
=
f_\theta(X_m,S)+\kappa(\mathbf{E}),
\qquad m\in L.
\label{eq:common_sequence_offset}
\end{equation}
Here, $f_\theta$ is the model score defined in the methodology;
$f_\theta^{+}$ is introduced only for this comparison. This construction
does not assume a decomposition of the model score.

Because $\kappa(\mathbf{E})$ is identical for all candidates in $L$,
it cancels from every pairwise difference:
\begin{equation}
\begin{aligned}
f_\theta^{+}(X_m,S)-f_\theta^{+}(X_n,S)
&=
\bigl[f_\theta(X_m,S)+\kappa(\mathbf{E})\bigr]
\\
&\quad-
\bigl[f_\theta(X_n,S)+\kappa(\mathbf{E})\bigr]
\\
&=
f_\theta(X_m,S)-f_\theta(X_n,S).
\end{aligned}
\label{eq:sequence_cancellation}
\end{equation}
Thus, the added term preserves all pairwise orderings and ties.

The ListNet objective is also unchanged. Let $\widehat{p}_m$ and
$\widehat{p}_m^{+}$ denote the predicted list probabilities obtained
from $f_\theta$ and $f_\theta^{+}$, respectively. For the ListNet
temperature $\tau>0$,
\begin{equation}
\begin{aligned}
\widehat{p}_m^{+}
&=
\frac{\exp\!\left(f_\theta^{+}(X_m,S)/\tau\right)}
{\sum_{n\in L}\exp\!\left(f_\theta^{+}(X_n,S)/\tau\right)}
\\
&=
\frac{
\exp\!\left(\kappa(\mathbf{E})/\tau\right)
\exp\!\left(f_\theta(X_m,S)/\tau\right)}
{
\exp\!\left(\kappa(\mathbf{E})/\tau\right)
\sum_{n\in L}\exp\!\left(f_\theta(X_n,S)/\tau\right)}
\\
&=
\widehat{p}_m.
\end{aligned}
\label{eq:sequence_softmax_invariance}
\end{equation}
The target probabilities $p_m$, determined by the within-list RMSD
ranks, remain fixed. Consequently,
\begin{equation}
\mathcal{L}_{\mathrm{rank}}^{+}
=
-\sum_{m\in L}p_m\log\widehat{p}_m^{+}
=
-\sum_{m\in L}p_m\log\widehat{p}_m
=
\mathcal{L}_{\mathrm{rank}},
\label{eq:sequence_loss_invariance}
\end{equation}
where the superscript $+$ identifies the loss evaluated using the
auxiliary scores. For the explicit sequence embedding input to affect
ranking, its contribution must therefore depend on the candidate
and change relative scores.

\paragraph{Sequence conditioning can reduce ranking risk through sequence--geometry interaction.} The above analysis motivates viewing sequence conditioning as \emph{sequence--geometry interaction}, rather than as an independent sequence bias. Across RNA targets, let \(S\) denote the target sequence, let \(\mathbf E=\psi(S)\) denote its contextual embedding produced by a fixed pretrained RNA encoder, let \(\mathbf U_L=(\mathbf U_m)_{m\in L}\) denote the geometric representations of the candidates in a list, and let \(J\in\{1,\ldots,|L|\}\) be the categorical top-one random variable induced by the ListNet target distribution, with \(P(J=m\mid\mathbf y)=p_m\), where \(\mathbf y\) denotes the ground-truth candidate qualities. For an information set \(Z\), define the Bayes-optimal risk under the top-one ListNet logarithmic loss as
\begin{equation}
R^{*}(Z)
=
\inf_{q(\cdot\mid Z)}
\mathbb{E}\!\left[-\log q(J\mid Z)\right],
\label{eq:bayes_optimal_risk}
\end{equation}
where the infimum is taken over all conditional predictive distributions \(q(J\mid Z)\). Under logarithmic loss, the optimum is attained by the true conditional distribution \(q^{*}(J\mid Z)=P(J\mid Z)\), yielding \(R^{*}(Z)=H(J\mid Z)\)~\citep{jiao2015justification}. Therefore, using geometry alone gives
\begin{equation}
R_{\mathrm{geo}}^{*}
=
R^{*}(\mathbf U_L)
=
H(J\mid\mathbf U_L),
\end{equation}
whereas additionally conditioning on the target sequence gives
\begin{equation}
R_{\mathrm{geo+seq}}^{*}
=
R^{*}(\mathbf U_L,S)
=
H(J\mid\mathbf U_L,S).
\end{equation}
Consequently, the reduction in Bayes-optimal ranking risk due to sequence conditioning is
\begin{equation}
\Delta R_{\mathrm{seq}}
=
R_{\mathrm{geo}}^{*}-R_{\mathrm{geo+seq}}^{*}
=
H(J\mid\mathbf U_L)-H(J\mid\mathbf U_L,S)
=
I(J;S\mid\mathbf U_L)
\ge 0.
\label{eq:sequence_risk_reduction}
\end{equation}
For the contextual embedding actually supplied to the model,
\begin{equation}
R_{\mathrm{geo+emb}}^{*}
=
R^{*}(\mathbf U_L,\mathbf E)
=
H(J\mid\mathbf U_L,\mathbf E),
\end{equation}
and the corresponding risk reduction is
\begin{equation}
\Delta R_{\mathrm{emb}}
=
R_{\mathrm{geo}}^{*}-R_{\mathrm{geo+emb}}^{*}
=
I(J;\mathbf E\mid\mathbf U_L)
\ge 0.
\label{eq:embedding_risk_reduction}
\end{equation}
Since \(\mathbf E=\psi(S)\) is a deterministic function of the sequence, the conditional data-processing inequality gives
\begin{equation}
I(J;\mathbf E\mid\mathbf U_L)
\le
I(J;S\mid\mathbf U_L).
\label{eq:embedding_data_processing}
\end{equation}
Thus, contextual embeddings cannot contain more ranking-relevant information than the underlying sequence, but they can still reduce the Bayes-optimal ranking risk whenever \(I(J;\mathbf E\mid\mathbf U_L)>0\). A strict reduction from full sequence conditioning occurs whenever \(I(J;S\mid\mathbf U_L)>0\), meaning that the RNA sequence contains information about candidate ranking that is not already determined by geometry alone.

The proposed benefit of pretraining is more specific. RiNALMo has
already learned contextual sequence dependencies from a large corpus,
making these features available to the evaluator without learning the
sequence encoder from ranked structures. General analyses of
self-supervised learning show that reconstructing information already
present in the input can nevertheless yield representations that reduce
downstream approximation error and labelled-data requirements under
appropriate statistical assumptions \citep{lee2021predicting}.
For SIRGE, the biological basis for transfer is the relationship between sequence dependencies and structural constraints. RiNALMo's transfer
after fine-tuning to secondary-structure prediction supports the
relevance of its representations to structural tasks
\citep{penic2025rinalmo}. Our claim is that exposing the geometric
evaluator to these pretrained features helps it learn compatibility
between sequence context and candidate structure, improving ranking
and generalisation to unseen RNAs. This is an empirical claim about
learning with finite supervision; the exact results above establish
the information and ranking properties within which that claim is
interpreted.
% We provide a derivation of Eq.~\ref{eq:conditional_information}. Let
% $\eta(\mathbf{z})=\Pr(Y_{mn}=1\mid Z_{mn}=\mathbf{z})$. For a predicted
% probability $p$, the conditional binary log loss is
% %
% \begin{equation}
%     -\eta(\mathbf{z})\log p
%     -
%     \bigl(1-\eta(\mathbf{z})\bigr)\log(1-p).
% \end{equation}
% %
% This quantity is minimised by $p=\eta(\mathbf{z})$. The Bayes-optimal
% geometry-only risk is therefore $H(Y_{mn}\mid Z_{mn})$. Conditioning additionally
% on the sequence representation $E$ gives the optimal risk
% $H(Y_{mn}\mid Z_{mn},E)$. Their difference is
% %
% \begin{equation}
%     \mathcal{R}_{\mathrm{geo}}^\star
%     -
%     \mathcal{R}_{\mathrm{cond}}^\star
%     =
%     I(Y_{mn};E\mid Z_{mn})
%     \geq 0.
% \end{equation}
% %
% Although $E$ is shared by candidates of the same RNA, it can change how their
% geometries are interpreted. The inequality is strict whenever the sequence
% representation contains ordering information that cannot be recovered from the
% geometric representation alone.

\subsection{Rigid-motion invariance}
\label{app:rigid_invariance}

Consider a global rigid transformation specified by a rotation
$\mathbf{Q}\in\mathrm{SO}(3)$ and translation
$\mathbf{t}\in\mathbb{R}^3$. Let $X_m'$ denote the transformed candidate, with
\begin{equation}
\mathbf{x}_{m,a}'
=
\mathbf{Q}\mathbf{x}_{m,a}+\mathbf{t}.
\label{eq:transformed_coordinates}
\end{equation}
The atom features $\mathbf{c}_{m,a}$ and nucleotide assignments $\rho(a)$ are
unchanged. Equivariance of the atom encoder gives
\begin{equation}
\mathbf{a}_{m,a}^{(0)\prime}
=
\mathbf{a}_{m,a}^{(0)},
\qquad
\mathbf{a}_{m,a}^{(1)\prime}
=
\mathbf{Q}\mathbf{a}_{m,a}^{(1)},
\label{eq:transformed_atom_features}
\end{equation}
so scalar channels remain fixed while vector channels rotate with the
candidate. Because each nucleotide frame is constructed from its atomic
coordinates, it co-transforms as
\begin{equation}
\mathbf{R}_{m,i}'
=
\mathbf{Q}\mathbf{R}_{m,i},
\qquad
\mathbf{o}_{m,i}'
=
\mathbf{Q}\mathbf{o}_{m,i}+\mathbf{t}.
\label{eq:transformed_frame}
\end{equation}
Using $\mathbf{Q}^{\top}\mathbf{Q}=\mathbf{I}$, the two local-frame projections
in \eqref{eq:local_projection} satisfy
\begin{align}
(\mathbf{R}_{m,i}')^{\top}\mathbf{a}_{m,a}^{(1)\prime}
&=
\mathbf{R}_{m,i}^{\top}\mathbf{a}_{m,a}^{(1)}, \\
(\mathbf{R}_{m,i}')^{\top}
\left(\mathbf{x}_{m,a}'-\mathbf{o}_{m,i}'\right)
&=
\mathbf{R}_{m,i}^{\top}
\left(\mathbf{x}_{m,a}-\mathbf{o}_{m,i}\right).
\end{align}
Together with the invariant scalar channels, these identities give
$\mathbf{d}_{m,a}'=\mathbf{d}_{m,a}$. Attention pooling therefore produces the
same geometric nucleotide representation,
$\widetilde{\mathbf{u}}_{m,i}'=\widetilde{\mathbf{u}}_{m,i}$.

The nucleotide identity $\mathbf{r}_i$ and RiNALMo embedding $\mathbf{e}_i$
depend only on the unchanged sequence $S$. Consequently,
$\mathbf{u}_{m,i}'=\mathbf{u}_{m,i}$ and
$\mathbf{h}_{m,i}^{\mathrm{cond}\prime}
=\mathbf{h}_{m,i}^{\mathrm{cond}}$.

It remains to consider geometric reasoning. Any learned point with local
coordinates $\mathbf{z}\in\mathbb{R}^3$ is mapped into global coordinates as
$\mathbf{R}_{m,i}\mathbf{z}+\mathbf{o}_{m,i}$. Under the rigid transformation,
this point becomes
\begin{equation}
\mathbf{R}_{m,i}'\mathbf{z}+\mathbf{o}_{m,i}'
=
\mathbf{Q}
\left(
\mathbf{R}_{m,i}\mathbf{z}+\mathbf{o}_{m,i}
\right)
+\mathbf{t}.
\end{equation}
Thus, differences between points are rotated by $\mathbf{Q}$, and their
Euclidean distances are unchanged. The scalar attention terms, point-distance
terms, and attention weights in every IPA block are therefore invariant.
Because the corresponding value projections are computed from invariant
nucleotide features, all IPA outputs remain invariant.

Finally, nucleotide pooling and the atom residual path operate only on
invariant features. Their outputs, and therefore their sum, are invariant,
yielding
\begin{equation}
f_\theta(X_m',S)=f_\theta(X_m,S).
\label{eq:rigid_motion_invariance}
\end{equation}

\subsection{Listwise objective and homogeneous batches}
\label{app:ranking_objective}

Following HomRank \citep{li2025homrank}, each training batch contains candidates
from a single RNA. We sample lists $\mathcal{L}$ from this homogeneous candidate
pool and optimise the top-one ListNet objective \citep{cao2007learning}. For each
candidate $m\in\mathcal{L}$, we convert its ascending RMSD rank into a relevance
score
\begin{equation}
q_m
=
\frac{1}{\operatorname{rank}_{\delta}(d_m)},
\end{equation}
where $\operatorname{rank}_{\delta}$ assigns the same rank to candidates whose
RMSDs differ by at most $\delta=0.1\,\text{\AA}$. We then define target and
predicted list distributions:
\begin{equation}
p_m
=
\frac{\exp(q_m/\tau)}
{\sum_{n\in\mathcal{L}}\exp(q_n/\tau)},
\qquad
\widehat{p}_m
=
\frac{\exp(f_\theta(X_m,S)/\tau)}
{\sum_{n\in\mathcal{L}}\exp(f_\theta(X_n,S)/\tau)},
\end{equation}
with temperature $\tau=0.1$. The loss is the cross-entropy between these
distributions:
\begin{equation}
\mathcal{L}_{\mathrm{rank}}
=
-\sum_{m\in\mathcal{L}}
p_m\log\widehat{p}_m.
\end{equation}
At this temperature, the target distribution is sharply concentrated on the
candidate or tied candidates with the highest relevance. The objective therefore
primarily supervises best-candidate selection rather than the complete permutation
of the list, while using the remaining candidates as competing alternatives. We
average the loss over the lists sampled from each homogeneous batch. This setup
ensures that every comparison is made between alternative conformations of the
same RNA.
\section{Architecture and Implementation Details}
\label{app:architecture}

\paragraph{Atomic encoder.}
We represent each candidate as a graph of heavy atoms. Node features encode the elements C, N, O, P, and S, and each atom is connected to its 32 nearest neighbours. The atom encoder contains three Equiformer blocks with four attention heads and spherical harmonics up to $\ell_{\max}=2$. Its hidden representation comprises $32$ scalar, $8$ vector, and $4$ second-order channels. Interatomic distances are expanded with $16$ radial basis functions over a range of $30$\,\AA.

\paragraph{Nucleotide projection and sequence conditioning.}
We place the origin of each nucleotide frame at C4$'$. The frame axes are constructed from P, C4$'$, and the glycosidic nitrogen N9 for purines or N1 for pyrimidines. Scalar atom features, vector features rotated into this frame, and local atom coordinates are pooled with four-head attention to produce a $64$-dimensional nucleotide representation. We extract the $1280$-dimensional representation of each nucleotide from RiNALMo's final layer, keep RiNALMo frozen, and apply layer normalisation followed by a learned $1280\!\rightarrow\!64$ projection. The projected embedding is added to the corresponding structure-derived representation before geometric reasoning. The projection is initialised to zero, so training begins from the geometry-only model. 

\paragraph{Nucleotide reasoning and readout.}
The conditioned representations pass through four dense IPA blocks, so every nucleotide can attend to every other nucleotide in the same candidate. Each block uses four heads, with $16$ scalar channels and four learned points per head. Coordinates are scaled by $0.1$ before point attention, converting their units from \AA{} to nm. A learned-query attention pool then produces a $64$-dimensional candidate representation, which is mapped to a scalar by a $64\!\rightarrow\!256\!\rightarrow\!1$ MLP with an ELU activation. In parallel, the scalar atom features are mean-pooled and passed through a second MLP of the same form. Its final layer is initialised to zero, and its output is added to the nucleotide-level score as the atom residual connection.
\section{Dataset Construction and Split Integrity}
\label{app:dataset}

\paragraph{Data sources and candidate generation.}
We use the dataset introduced by HomRank \citep{li2025homrank}. It contains $190$ non-redundant, single-chain RNA targets: $151$ structures collected from the Protein Data Bank, $18$ targets from the ARES dataset, and $21$ RNA-Puzzle targets. For each of the $151$ newly collected targets, HomRank generated $1{,}000$ candidate conformations through molecular-dynamics simulations. The ARES targets also contain $1{,}000$ candidates per RNA. The RNA-Puzzle benchmark contains $451{,}629$ candidates generated through Rosetta FARFAR2 sampling. Candidate quality is measured by RMSD to the corresponding experimentally resolved structure.

\paragraph{Sequence-based partitioning.}
HomRank computes pairwise sequence similarities with the \texttt{Bio.Align} module and applies hierarchical clustering to the complete set of RNA targets. This produces three sequence-similarity clusters: CLS-0 with $146$ RNAs, CLS-1 with $21$ RNAs, and CLS-2 with $23$ RNAs. Fifteen RNA-Puzzle targets assigned to CLS-0 are excluded from training because RNA-Puzzles is an established benchmark. The remaining $131$ CLS-0 targets form our training set. We use CLS-2 for validation and checkpoint selection, and CLS-1 for final evaluation.

\begin{table}[t]
    \centering
    \caption{Dataset partitions used in our experiments. Candidate counts refer to the complete source benchmark before evaluation-time subsampling.}
    \label{tab:dataset_splits}
    \small
    \begin{tabular}{lrrr}
        \toprule
        Partition & RNA targets & Length range & Candidates \\
        \midrule
        Training (CLS-0)   & 131 & 22--144  & 131,000 \\
        Validation (CLS-2) & 23  & 12--49   & 23,000 \\
        Test (CLS-1)       & 21  & 61--257  & 152,917 \\
        \bottomrule
    \end{tabular}
\end{table}

\paragraph{Leakage controls.}
The split is performed at the RNA-target level rather than at the candidate level. Consequently, the sequence, native structure, and complete candidate ensemble associated with an RNA remain in a single partition. No candidates from a validation or test RNA appear during training. Moreover, training, validation, and test targets belong to distinct sequence-similarity clusters, reducing leakage between biologically related molecules. For evaluation, we use at most $2{,}000$ candidates per test RNA and apply the same candidate subsets to every method. Test labels do not contribute to gradient updates or checkpoint selection; all reported predictions use the checkpoint selected on CLS-2. RiNALMo remains frozen and receives only the RNA sequence, without candidate coordinates, RMSD labels, or native structures.
\section{Experimental Protocol}
\label{app:experimental_protocol}

\paragraph{Baselines.}
We compare SIRGE with six learned evaluators that cover complementary modelling approaches: the voxel-based RNA3DCNN \citep{li2018rna3dcnn}, the physics-aware graph network PAMNet \citep{zhang2023universal}, the frame-based lociPARSE model \citep{tarafder2024lociparse}, the hierarchical equivariant EquiRNA model \citep{li2025equirna}, the listwise HomRank model \citep{li2025homrank} and the canonical work for geometric RNA evaluation ARES \citep{townshend2021geometric}, as well as Rosetta \citep{watkins2020farfar2} . Each baseline retains its published architecture and learning objective.

\paragraph{Training and model selection.}
We retrain every learned method for five epochs on the same CLS-0 training set. Model-specific optimisation settings follow the corresponding published implementations. SIRGE is trained with AdamW using a learning rate of $3\times10^{-3}$, weight decay of $10^{-4}$, gradient clipping at $5.0$, a $100$-step linear warm-up, and cosine learning-rate decay. Each homogeneous batch contains $16$ candidates from one RNA. We sample $32$ lists of length $16$ for the ListNet objective and accumulate gradients across eight batches. For every method, the best checkpoint is selected using its validation criterion on CLS-2 and subsequently evaluated on CLS-1 without further optimisation.

\paragraph{Evaluation metrics.}
All predictions are oriented as goodness scores, so larger values indicate better candidate structures. Kendall alignment is computed from the full candidate pool within each RNA as $-\tau(f_\theta,\mathrm{RMSD})$, making $+1$ perfect agreement. Top-1 and Top-3 ranking accuracy are evaluated on $4{,}000$ uniformly sampled lists of ten candidates per RNA. Top-1 measures whether the highest-scoring candidate has the lowest RMSD. Top-3 measures whether the three lowest-RMSD candidates are placed in the correct relative order; RMSD differences below $0.1$\,\AA{} are treated as ties. We use identical sampled lists for every method and macro-average each metric across RNA targets.

\paragraph{Reproducibility and hardware.}
We evaluate each learned method over three independent training runs and report the mean and standard deviation. All experiments are conducted on a system with four NVIDIA RTX PRO 6000 Blackwell GPUs, each with $98$\,GB of memory.
\section{Complete Quantitative Results}
\label{app:complete_results}
% \begin{table}[t]
% \centering
% \caption{Top-1 retrieval (\%) on Test-CLS-1. Higher is better.}
% \label{tab:complete_top1}
% \begin{tabular}{@{}lcccc@{}}
% \toprule
% Model & Global & D1 & D2 & D3 \\ 
% \midrule
% Random & 10.0 & 10.0 & 10.0 & 10.0 \\
% Rosetta & 12.4 & 13.1 & 8.7 & 11.2 \\
% ARES & 22.0 $\pm$ 1.4 & 36.2 $\pm$ 2.3 & 21.9 $\pm$ 1.3 & 15.4 $\pm$ 0.6 \\
% PAMNet & 16.7 $\pm$ 0.1 & 30.9 $\pm$ 0.4 & 14.3 $\pm$ 0.6 & 9.6 $\pm$ 0.4 \\
% RNA3DCNN & 16.3 $\pm$ 0.7 & 31.7 $\pm$ 1.7 & 19.2 $\pm$ 3.1 & 12.7 $\pm$ 1.7 \\
% EquiRNA & 16.5 $\pm$ 1.6 & 27.6 $\pm$ 8.4 & 18.3 $\pm$ 4.3 & 13.1 $\pm$ 0.5 \\
% lociPARSE & 20.5 $\pm$ 1.2 & 40.2 $\pm$ 1.4 & 22.0 $\pm$ 2.1 & 14.3 $\pm$ 3.1 \\
% HomRank & 23.2 $\pm$ 0.8 & 42.5 $\pm$ 5.5 & 33.8 $\pm$ 0.4 & 19.2 $\pm$ 2.4 \\
% \textsc{Sirge}-SO & 25.4 $\pm$ 2.1 & 45.4 $\pm$ 2.5 & 35.3 $\pm$ 2.9 & 23.6 $\pm$ 3.5 \\
% \textsc{Sirge} & \textbf{31.7 $\pm$ 1.2} & \textbf{50.1 $\pm$ 2.7} & \textbf{43.9 $\pm$ 1.8} & \textbf{33.0 $\pm$ 1.5} \\
% \bottomrule
% \end{tabular}
% % }
% \end{table}

\begin{table}[t]
\centering
\caption{
Top-1 retrieval (\%) on \mbox{Test-CLS-1}. Results are reported as mean
$\pm$ std over three random seeds. Higher is better.
}
\label{tab:complete_top1}

\scriptsize
\setlength{\tabcolsep}{4pt}
\renewcommand{\arraystretch}{1.18}

\resizebox{\linewidth}{!}{
\begin{tabular}{@{}lcccc@{}}
\toprule
Model
& Global
& D1 ($2$--$5$\,\AA)
& D2 ($5$--$10$\,\AA)
& D3 ($10$--$15$\,\AA) \\
\midrule

Random
& 10.0
& 10.0
& 10.0
& 10.0 \\

Rosetta
& 12.4
& 13.1
& 8.7
& 11.2 \\[2pt]

ARES
& $22.0 \pm 1.4$
& $36.2 \pm 2.3$
& $21.9 \pm 1.3$
& $15.4 \pm 0.6$ \\

PAMNet
& $16.7 \pm 0.1$
& $30.9 \pm 0.4$
& $14.3 \pm 0.6$
& $9.6 \pm 0.4$ \\

RNA3DCNN
& $16.3 \pm 0.7$
& $31.7 \pm 1.7$
& $19.2 \pm 3.1$
& $12.7 \pm 1.7$ \\

EquiRNA
& $16.5 \pm 1.6$
& $27.6 \pm 8.4$
& $18.3 \pm 4.3$
& $13.1 \pm 0.5$ \\

lociPARSE
& $20.5 \pm 1.2$
& $40.2 \pm 1.4$
& $22.0 \pm 2.1$
& $14.3 \pm 3.1$ \\[2pt]

HomRank
& $23.2 \pm 0.8$
& $42.5 \pm 5.5$
& $33.8 \pm 0.4$
& $19.2 \pm 2.4$ \\

\midrule

\rowcolor{gray!20}
\textbf{SIRGE-SO}
& $25.4 \pm 2.1$
& $45.4 \pm 2.5$
& $35.3 \pm 2.9$
& $23.6 \pm 3.5$ \\

\rowcolor{gray!20}
\textbf{SIRGE}
& $\mathbf{31.7} \pm 1.2$
& $\mathbf{50.1} \pm 2.7$
& $\mathbf{43.9} \pm 1.8$
& $\mathbf{33.0} \pm 1.5$ \\

\bottomrule
\end{tabular}
}
\end{table}
% \begin{table}[t]
% \centering
% \caption{Top-3 ranking (\%) on Test-CLS-1. Higher is better.}
% \label{tab:complete_top3}
% \begin{tabular}{@{}lcccc@{}}
% \toprule
% Model & Global & D1 & D2 & D3 \\ 
% \midrule
% Random & 19.7 & 19.7 & 19.7 & 19.7 \\
% Rosetta & 20.2 & 19.1 & 18.7 & 17.3 \\
% ARES & 26.2 $\pm$ 0.8 & 33.0 $\pm$ 1.8 & 27.8 $\pm$ 0.6 & 24.5 $\pm$ 0.6 \\
% PAMNet & 22.3 $\pm$ 0.2 & 29.5 $\pm$ 2.4 & 20.4 $\pm$ 0.7 & 20.3 $\pm$ 1.6 \\
% RNA3DCNN & 22.3 $\pm$ 0.3 & 28.0 $\pm$ 1.6 & 23.9 $\pm$ 1.8 & 21.1 $\pm$ 1.3 \\
% EquiRNA & 23.0 $\pm$ 1.2 & 28.8 $\pm$ 5.1 & 25.5 $\pm$ 3.7 & 22.0 $\pm$ 0.4 \\
% lociPARSE & 25.9 $\pm$ 0.9 & 35.0 $\pm$ 0.9 & 27.2 $\pm$ 0.8 & 22.6 $\pm$ 2.5 \\
% HomRank & 26.8 $\pm$ 0.1 & 36.7 $\pm$ 1.9 & 33.7 $\pm$ 0.3 & 25.1 $\pm$ 1.1 \\
% \textsc{Sirge}-SO & 29.5 $\pm$ 2.1 & 38.5 $\pm$ 1.7 & 34.3 $\pm$ 2.9 & 26.1 $\pm$ 2.0 \\
% \textsc{Sirge} & \textbf{31.3 $\pm$ 0.6} & \textbf{41.5 $\pm$ 0.4} & \textbf{38.8 $\pm$ 2.4} & \textbf{31.6 $\pm$ 0.4} \\
% \bottomrule
% \end{tabular}
% % }
% \end{table}

\begin{table}[t]
\centering
\caption{
Top-3 ranking (\%) on \mbox{Test-CLS-1}. Results are reported as mean
$\pm$ std over three random seeds. Higher is better.
}
\label{tab:complete_top3}

\scriptsize
\setlength{\tabcolsep}{4pt}
\renewcommand{\arraystretch}{1.18}

\resizebox{\linewidth}{!}{
\begin{tabular}{@{}lcccc@{}}
\toprule
Model
& Global
& D1 ($2$--$5$\,\AA)
& D2 ($5$--$10$\,\AA)
& D3 ($10$--$15$\,\AA) \\
\midrule

Random
& 19.7
& 19.7
& 19.7
& 19.7 \\

Rosetta
& 20.2
& 19.1
& 18.7
& 17.3 \\[2pt]

ARES
& $26.2 \pm 0.8$
& $33.0 \pm 1.8$
& $27.8 \pm 0.6$
& $24.5 \pm 0.6$ \\

PAMNet
& $22.3 \pm 0.2$
& $29.5 \pm 2.4$
& $20.4 \pm 0.7$
& $20.3 \pm 1.6$ \\

RNA3DCNN
& $22.3 \pm 0.3$
& $28.0 \pm 1.6$
& $23.9 \pm 1.8$
& $21.1 \pm 1.3$ \\

EquiRNA
& $23.0 \pm 1.2$
& $28.8 \pm 5.1$
& $25.5 \pm 3.7$
& $22.0 \pm 0.4$ \\

lociPARSE
& $25.9 \pm 0.9$
& $35.0 \pm 0.9$
& $27.2 \pm 0.8$
& $22.6 \pm 2.5$ \\[2pt]

HomRank
& $26.8 \pm 0.1$
& $36.7 \pm 1.9$
& $33.7 \pm 0.3$
& $25.1 \pm 1.1$ \\

\midrule

\rowcolor{gray!20}
\textbf{SIRGE-SO}
& $29.5 \pm 2.1$
& $38.5 \pm 1.7$
& $34.3 \pm 2.9$
& $26.1 \pm 2.0$ \\

\rowcolor{gray!20}
\textbf{SIRGE}
& $\mathbf{31.3} \pm 0.6$
& $\mathbf{41.5} \pm 0.4$
& $\mathbf{38.8} \pm 2.4$
& $\mathbf{31.6} \pm 0.4$ \\

\bottomrule
\end{tabular}
}
\end{table}
% \begin{table}[t]
% \centering
% \caption{Spearman rank alignment, $\rho_{\mathrm{align}}=\rho(f_\theta,-d)$ on Test-CLS-1. Higher is better.}
% \label{tab:complete_spearman}
% \begin{tabular}{@{}lcccc@{}}
% \toprule
% Model & Global & D1 & D2 & D3 \\ 
% \midrule
% Rosetta & 0.021 & 0.007 & -0.033 & 0.011 \\
% ARES & 0.189 $\pm$ 0.018 & 0.285 $\pm$ 0.022 & 0.203 $\pm$ 0.010 & 0.084 $\pm$ 0.009 \\
% PAMNet & 0.097 $\pm$ 0.006 & 0.219 $\pm$ 0.022 & 0.062 $\pm$ 0.011 & -0.046 $\pm$ 0.018 \\
% RNA3DCNN & 0.107 $\pm$ 0.025 & 0.183 $\pm$ 0.009 & 0.112 $\pm$ 0.030 & 0.027 $\pm$ 0.009 \\
% EquiRNA & 0.108 $\pm$ 0.026 & 0.224 $\pm$ 0.071 & 0.116 $\pm$ 0.031 & 0.078 $\pm$ 0.020 \\
% lociPARSE & 0.194 $\pm$ 0.018 & 0.343 $\pm$ 0.017 & 0.185 $\pm$ 0.025 & 0.079 $\pm$ 0.033 \\
% HomRank & 0.212 $\pm$ 0.020 & 0.299 $\pm$ 0.034 & 0.275 $\pm$ 0.024 & 0.146 $\pm$ 0.016 \\
% \textsc{Sirge}-SO & 0.261 $\pm$ 0.030 & 0.498 $\pm$ 0.078 & 0.344 $\pm$ 0.031 & 0.184 $\pm$ 0.020 \\
% \textsc{Sirge} & \textbf{0.350 $\pm$ 0.033} & \textbf{0.556 $\pm$ 0.005} & \textbf{0.442 $\pm$ 0.009} & \textbf{0.266 $\pm$ 0.040} \\
% \bottomrule
% \end{tabular}
% % }
% \end{table}

\begin{table}[t]
\centering
\caption{
Spearman rank alignment,
$\rho_{\mathrm{align}}=\rho(f_\theta,-d)$, on \mbox{Test-CLS-1}.
Results are reported as mean $\pm$ std over three random seeds.
Higher is better.
}
\label{tab:complete_spearman}

\scriptsize
\setlength{\tabcolsep}{4pt}
\renewcommand{\arraystretch}{1.18}

\resizebox{\linewidth}{!}{
\begin{tabular}{@{}lcccc@{}}
\toprule
Model
& Global
& D1 ($2$--$5$\,\AA)
& D2 ($5$--$10$\,\AA)
& D3 ($10$--$15$\,\AA) \\
\midrule

Rosetta
& 0.021
& 0.007
& -0.033
& 0.011 \\[2pt]

ARES
& $0.189 \pm 0.018$
& $0.285 \pm 0.022$
& $0.203 \pm 0.010$
& $0.084 \pm 0.009$ \\

PAMNet
& $0.097 \pm 0.006$
& $0.219 \pm 0.022$
& $0.062 \pm 0.011$
& $-0.046 \pm 0.018$ \\

RNA3DCNN
& $0.107 \pm 0.025$
& $0.183 \pm 0.009$
& $0.112 \pm 0.030$
& $0.027 \pm 0.009$ \\

EquiRNA
& $0.108 \pm 0.026$
& $0.224 \pm 0.071$
& $0.116 \pm 0.031$
& $0.078 \pm 0.020$ \\

lociPARSE
& $0.194 \pm 0.018$
& $0.343 \pm 0.017$
& $0.185 \pm 0.025$
& $0.079 \pm 0.033$ \\[2pt]

HomRank
& $0.212 \pm 0.020$
& $0.299 \pm 0.034$
& $0.275 \pm 0.024$
& $0.146 \pm 0.016$ \\

\midrule

\rowcolor{gray!20}
\textbf{SIRGE-SO}
& $0.261 \pm 0.030$
& $0.498 \pm 0.078$
& $0.344 \pm 0.031$
& $0.184 \pm 0.020$ \\

\rowcolor{gray!20}
\textbf{SIRGE}
& $\mathbf{0.350} \pm 0.033$
& $\mathbf{0.556} \pm 0.005$
& $\mathbf{0.442} \pm 0.009$
& $\mathbf{0.266} \pm 0.040$ \\

\bottomrule
\end{tabular}
}
\end{table}

Tables~\ref{tab:complete_top1}--\ref{tab:complete_spearman} complement the Kendall--$\tau$ results in Table~\ref{tab:main_results} with three additional measures of candidate selection and ranking. Top-1 retrieval measures whether the evaluator selects the lowest-RMSD candidate from a sampled list of ten structures. Top-3 ranking measures whether it correctly orders the three best candidates, while Spearman alignment measures monotonic agreement with the RMSD ordering. We report results globally and for D1 ($2$--$5$\,\AA), D2 ($5$--$10$\,\AA), and D3 ($10$--$15$\,\AA) lists, where each list is assigned to the band containing its lowest-RMSD candidate.

Results are first averaged within each RNA and then across RNAs, so every molecule contributes equally. The neural baselines report mean $\pm$ standard deviation over three training seeds. Random denotes the expected performance of a uniformly random ordering.

\textsc{Sirge} achieves the strongest result for every metric, both globally and across all three quality regimes. \textsc{Sirge}-SO already improves over existing evaluators, demonstrating the strength of the hierarchical geometric architecture. Sequence conditioning provides a further improvement in every setting, with particularly clear gains in D2 and D3. These results reinforce the Kendall-alignment findings in Table~\ref{tab:main_results}: contextual sequence representations improve both overall rank agreement and the candidate-selection decisions that matter most in practice.

%%%%%%%%%%%%%%%%%%%%%%%%%%%%%%%%%%%%%%%%%%%%%%%%%%%%%%%%%%%%

\end{document}